\documentclass[letter,twocolumn]{jpsj3}
\usepackage{txfonts}
\usepackage{graphicx,bm,color,braket}

\title{Charging due to Pair-Potential Gradient in Vortex of Type-II Superconductors}

\author{Marie Ohuchi, Hikaru Ueki, and Takafumi Kita}
\inst{Department of Physics, Hokkaido University, Sapporo 060-0810, Japan} 

\abst{
Besides the magnetic Lorentz force familiar from the Hall effect in metals and semiconductors, 
there exists a mechanism for charging peculiar to superconductors that is caused by the pair-potential gradient (PPG).
We incorporate it in the augmented quasiclassical equations of superconductivity with the Lorentz force to
study charging of an isolated vortex in an equilibrium $s$-wave type-II superconductor.
It is found that the PPG mechanism gives rise to charging concentrated within the core
whose magnitude at the core center can be 10 to $10^2$ times larger than that caused by the Lorentz force.
Our detailed calculations on the spatial, temperature, and magnetic-penetration-depth dependences
of the vortex-core charge reveal that the  PPG mechanism contributes dominantly to the core charging of the isolated vortex
over a wide parameter range. 
The two mechanisms are also found to work additively at the core center for the present model with an isotropic Fermi surface.
}

\begin{document}
\maketitle

Superconductors are characterized by the emergence of the pair potential,
whose amplitude is reduced to zero towards the edges and into the vortex cores over the coherence length $\xi$.
This spatial variation in the pair potential is a cause of the charging 
characteristic of superconductors.
The relevant term was obtained more than two decades ago by Kopnin\cite{Kopnin} in his derivation of
kinetic equations for clean superconductors as a next-to-leading-order contribution
in the expansion in terms of  the quasiclassical parameter $\delta\equiv 1/k_{\rm F}\xi_0\ll 1$,
where $k_{\rm F}$ is the Fermi wavenumber and $\xi_0\equiv \hbar v_{\rm F}/\Delta_0$ denotes the coherence length 
defined in terms of the zero-temperature  energy gap $\Delta_0$ in a zero magnetic field.

This force was also present in a later gauge-invariant derivation of the augmented quasiclassical equations
for superconductivity with the Lorentz force\cite{Kita01} but was omitted inappropriately at the final stage.
It was recovered later in the augmented quasiclassical equations  by Arahata and Kato,\cite{AK}
with which they performed a numerical study 
on the nonequilibrium Hall effect with a moving isolated vortex. 
Its possible relevance to the charging in superconductors was pointed out by Hoshino and Kato.\cite{Hoshino}
To the best of our knowledge, however, no detailed quantitative calculations have been performed yet
on how {\em this pair-potential-gradient (PPG) force} affects charging in superconductors.
Here, we present a numerical study on the charging of an isolated vortex 
due to the PPG force in comparison with the Lorentz force based on the augmented quasiclassical equations
for superconductivity.\cite{Kita01,Kita09,Ueki1,Kohno1,Kohno2}
Note that the PPG force constitutes a mechanism independent of the one proposed by 
Khomskii and Freimuth\cite{Khomskii} and Feigel'man {\em et al.}\cite{Feigel'man}
in that it acts even when no change in the chemical potential arises between the normal and (homogeneous) superconducting states.

The augmented quasiclassical equations  include the Eilenberger equations\cite{Eilenberger,SR,LO86,KopninText,KitaText}
as the leading-order equations in the quasiclassical parameter $\delta$.
The Eilenberger equations are now regarded as
one of the most powerful methods 
for studying inhomogeneous and/or nonequilibrium superconductors microscopically\cite{Eliashberg,KP,SR,LO86,Klein,SM,Ichioka96,Eschrig,KopninText,KitaText}. 
However, the standard equations cannot describe either charging or the Hall effect in superconductors,
the relevant terms for which emerge as the next-to-leading-order contribution in the gauge-invariant derivation
of the quasiclassical equations.\cite{Kita01}
The PPG force is also classified as the next-to-leading-order contribution\cite{Kopnin}
and can be incorporated naturally in the augmented quasiclassical equations.\cite{Kita01,AK}
Among their definite advantages over microscopic treatments\cite{Hayashi,MH,Chen,Knapp,MK} based on the Bogoliubov--de Gennes (BdG) equations\cite{CdGM,GennesText}
are that one can  (i) easily incorporate the Fermi-surface and gap anisotropies
that are crucial in determining the sign and magnitude of the Hall coefficient, (ii) trace distinct mechanisms for charging, 
and (iii) study equilibrium and nonequilibrium properties on an equal footing.
It is worth pointing out in this context that the augmented quasiclassical equations naturally reduce to the normal-state  Boltzmann equation 
with the Lorentz force when the pair potential is equal to zero.\cite{Kita01}
Moreover, they have successfully been transformed from the real-time Keldysh formalism \cite{Kita01,AK}
 into the Matsubara formalism\cite{Ueki1} to perform numerical calculations on the equilibrium 
properties and linear-response functions efficiently. 
Thus, the augmented quasiclassical equations are a powerful tool for describing equilibrium and nonequilibrium superconductors with $\delta\ll 1$,
at the expense of neglecting some quantum effects such as the Friedel oscillations that are predicted to  emerge as $\delta\rightarrow 1$
in the fully microscopic BdG approach.\cite{Hayashi,MH,Chen,Knapp,MK} 

For clean superconductors in equilibrium, 
the augmented quasiclassical equations 
with the PPG and Lorentz forces
are given in the Matsubara formalism by \cite{Kita01,AK,Ueki1}
\begin{eqnarray}
\bigl[ i \varepsilon_n \hat{\tau}_3 - \hat{\Delta}, \hat{g} \bigr] + i  \hbar {\bm v}_{\rm F} \cdot {\bm\partial} \hat{g} \notag  + \frac{i  \hbar}{2}e ({\bm v}_{\rm F} \times {\bm B}) \cdot \frac{\partial}{\partial {\bm p}_{\rm F}} \bigl\{ \hat{\tau}_3, \hat{g} \bigr\} \\
-\frac{i\hbar}{2}{\bm \partial}\hat{\Delta}\cdot\frac{\partial \hat{g}}{\partial {\bm p}_{\rm F}} -\frac{i\hbar}{2}\frac{\partial \hat{g}}{\partial {\bm p}_{\rm F}}\cdot{\bm \partial}\hat{\Delta} = \hat{0}. \label{LandK}
\end{eqnarray}
Here, $\hat{g}\!=\!\hat{g}(\varepsilon_n,{\bm p}_{\rm F},{\bm r})$ is the quasiclassical Green's function, $\varepsilon_n\!=\! (2n+1)\pi k_{\rm B}T$ is the fermion Matsubara energy $(n=0,\pm 1,\cdots)$ with $k_{\rm{B}}$ and $T$ denoting the Boltzmann constant and temperature, $\hat{\Delta}$ and  $\hat{\tau}_3$ are the $4\times 4$ gap matrix and third Pauli matrix in the Nambu space, respectively, ${\bm v}_{\rm F}$ and ${\bm p}_{\rm F}$ are the Fermi velocity and momentum, respectively, $e < 0$ is the electron charge, ${\bm B}$ denotes the magnetic-flux density, 
$\bm{\partial} $ denotes the gauge-invariant differential operator, $[ \hat{a}, \hat{b} ] \equiv \hat{a} \hat{b} - \hat{b} \hat{a}$, and $\{ \hat{a}, \hat{b} \} \equiv \hat{a} \hat{b} + \hat{b} \hat{a}$. 
The first and second terms in Eq.\ (\ref{LandK})  constitute the standard Eilenberger equations,
the third term denotes the Lorentz force, and the fourth and fifth terms represent the PPG force.
Matrices $\hat{g}$, $\hat{\Delta}$ and $\hat{\tau}_3$ can be written as \cite{KitaText}
\begin{align}
\hat{g}  \equiv
\begin{bmatrix}
\vspace{1mm}
\underline{g} & - i \underline{f} \\
- i \underline{\bar{f}} & - \underline{\bar{g}}
\end{bmatrix}, \hspace{3mm}
\hat{\Delta} \equiv
\begin{bmatrix}
\vspace{1mm}
\underline{0} & \underline{\Delta} \\
- \underline{\Delta^{\!*}} & \underline{0}
\end{bmatrix},\hspace{3mm}
\hat{\tau}_3 \equiv
\begin{bmatrix}
\vspace{1mm}
\underline{\sigma}_0 & \underline{0} \\
\underline{0} & -\underline{\sigma}_0
\end{bmatrix},
\end{align}
where the barred functions are defined generally by $\underline{\bar{g}} (\varepsilon_n, {\bm p}_{\rm F}, {\bm r}) \equiv \underline{g}^* (\varepsilon_n, - {\bm p}_{\rm F}, {\bm r})$,
and $\underline{\sigma}_0$ is the $2\times 2$ unit matrix. 
Operator ${\bm \partial}$ is given explicitly by
\begin{subequations}
\begin{align}
{\bm \partial} \equiv \left\{ \begin{array}{ll} {\bm \nabla} &  {\rm on} \  \underline{g} \ {\rm or} \ \bar{\underline{g}} \\
\displaystyle {\bm \nabla} - i \frac{2 e {\bm A}}{\hbar} &  {\rm on} \ \underline{f} \ {\rm or} \ \underline{\Delta} \\
\displaystyle  {\bm \nabla} + i \frac{2 e {\bm A}}{\hbar} & {\rm on} \ \bar{\underline{f}} \ {\rm or} \ \bar{\underline{\Delta}}
\end{array}\right. ,
\end{align} 
\end{subequations}
where ${\bm A}$ denotes the vector potential.

We consider the spin-singlet pairing without spin paramagnetism. 
Functions $\underline{g}$, $\underline{f}$ and $\underline{\Delta}$  can then be expressed as 
\begin{align}
\underline{g}  =
\begin{bmatrix}
\vspace{1mm}
g &  0 \\
0 &  g
\end{bmatrix}, \hspace{5mm}
\underline{f}  =
\begin{bmatrix}
\vspace{1mm}
0 &  -f \\
f &  0
\end{bmatrix}, \hspace{5mm}
\underline{\Delta}  =
\begin{bmatrix}
\vspace{1mm}
0 &  -\Delta \\
\Delta &  0
\end{bmatrix}.
\end{align}
As in Ref.\ \citen{Kita09}, we expand $g$ formally in $\delta$ as $g=g_0+g_1+\cdots$ and  $f=f_0+f_1+\cdots$, where
$g_0$ and $f_0$ are the solutions of the standard Eilenberger equations 
that satisfy the normalization condition $g_0\!=\!{\rm{sgn}}(\varepsilon_n)\bigl(1-f_0\bar{f}_0\bigr)^{1/2}$.\cite{KitaText,SR,KopninText}
The equation for $f_0$ is given by\cite{Kopnin,KitaText,SR,LO86}
\begin{subequations}
\label{Ei}
\begin{align}
\varepsilon_n f_0+\frac{1}{2}\hbar{\bm v}_{\rm{F}}\cdot \left({\bm \nabla} - i \frac{2 e {\bm A}}{\hbar}\right) f_0=\Delta g_0.
\label{Ei1}
\end{align}
On the other hand, the pair potential $\Delta$ and  the vector potential ${\bm A}$ are determined  by\cite{Kopnin,KitaText,SR,LO86}
\begin{align}
\Delta=\Gamma_0 \pi  k_{\rm{B}} T \sum_{n=-\infty}^{\infty}\langle f_0 \rangle_{\rm{F}}\label{Ei2},
\end{align}
\begin{align}
\bm{\nabla}\times\bm{\nabla}\times{\bm A}=-i2\pi e\mu_0N(0)k_{\rm{B}}T\sum_{n=-\infty}^{\infty}\langle {\bm v}_{\rm{F}}g_0 \rangle_{\rm{F}}\label{Ei3},
\end{align}
\end{subequations}
where $\Gamma_0\ll 1$ is the dimensionless coupling constant responsible for the Cooper pairing, $\langle\cdots\rangle_{\rm F}$ 
denotes the Fermi surface average normalized as $\langle 1\rangle_{\rm F}=1$,
 $\mu_0$ is the vacuum permeability, and $N(0)$ is the normal density of states per spin and unit volume at the Fermi energy.
Equation (\ref{Ei}) forms a set of self-consistent equations for $f_0$, $\Delta$, and ${\bm A}$.

The equations for $g_1$ and $f_1$ can be derived from Eq. (\ref{LandK}) as
\begin{subequations}
\begin{align}
&\Delta \bar{f}_1 - \Delta^{\!*} f_1 + \frac{i \hbar}{2}{\bm \partial}\Delta \cdot\frac{\partial \bar{f}_0}{\partial {\bm p}_{\rm F}} + \frac{i \hbar}{2}{\bm \partial}\Delta^{\!*} \cdot\frac{\partial f_0}{\partial {\bm p}_{\rm F}} \notag\\
& \hspace{23mm}+ \hbar{\bm v}_{\rm F}\cdot {\bm \nabla}g_1 + \hbar e ({\bm v}_{\rm F}\times{\bm B})\cdot \frac{\partial g_0}{\partial {\bm p}_{\rm F}} = 0, 
\end{align}
\begin{align}
&2\varepsilon_n f_1 - \Delta\bar{g}_1-\Delta g_1 - \frac{i \hbar}{2}{\bm \partial}\Delta \cdot\frac{\partial \bar{g}_0}{\partial {\bm p}_{\rm F}} + \frac{i \hbar}{2}{\bm \partial}\Delta \cdot\frac{\partial g_0}{\partial {\bm p}_{\rm F}} \notag\\
&\hspace{50mm}+ \hbar{\bm v}_{\rm F}\cdot {\bm \partial}f_1 = 0.
\end{align}
\end{subequations}
Taking their complex conjugates with ${\bm p}_{\rm F}\rightarrow-{\bm p}_{\rm F}$ and using $g_0 = \bar{g}_0$, we obtain
the four equations for $g_1+\bar{g}_1$, $g_1-\bar{g}_1$, $f_1$, and $\bar{f}_1$ as follows:
\begin{subequations}
\begin{align}
&\frac{\hbar}{2} {\bm v}_{\rm F}\cdot {\bm \nabla}(g_1-\bar{g}_1) + \hbar e ({\bm v}_{\rm F}\times{\bm B})\cdot \frac{\partial g_0}{\partial {\bm p}_{\rm F}} \notag\\
&\hspace{25mm}+ \frac{i \hbar}{2}{\bm \partial}\Delta \cdot\frac{\partial \bar{f}_0}{\partial {\bm p}_{\rm F}} + \frac{i \hbar}{2}{\bm \partial}\Delta^{\!*} \cdot\frac{\partial f_0}{\partial {\bm p}_{\rm F}}=0,\label{a}
\end{align}
\begin{align}
2\Delta \bar{f}_1 - 2\Delta^{\!*} f_1 + \hbar{\bm v}_{\rm F}\cdot {\bm \nabla}(g_1 +\bar{g}_1)  = 0, \label{b}
\end{align}
\begin{align}
2\varepsilon_n f_1  + \hbar{\bm v}_{\rm F}\cdot {\bm \partial}f_1 - \Delta\left(\bar{g}_1 + g_1\right) = 0,
\label{c}
\end{align}
\begin{align}
2\varepsilon_n \bar{f}_1 - \hbar{\bm v}_{\rm F}\cdot {\bm \partial}f_1 - \Delta^{\!*}\left(\bar{g}_1 + g_1\right) = 0. \label{d}
\end{align}
\end{subequations}
Equations (\ref{b}), (\ref{c}), and (\ref{d}) constitute linear closed equations without external sources.
We hence conclude $f_1 = 0$ and $g_1 = -\bar{g}_1$.
The substitution of this result into Eq.\ (\ref{a}) yields
\begin{equation}
{\bm v}_{\rm F} \cdot {\bm \nabla} g_1 = - e ({\bm v}_{\rm F} \times {\bm B}) \cdot \frac{\partial g_0}{\partial {\bm p}_{\rm F}} - \frac{i}{2} {\bm \partial}\Delta^{\!*} \cdot  \frac{\partial f_0}{\partial {\bm p}_{\rm F}} - \frac{i}{2} {\bm \partial} \Delta \cdot  \frac{\partial \bar f_0}{\partial {\bm p}_{\rm F}}. 
\label{nablag1}
\end{equation}
The charge density $\rho$ is obtained by \cite{KopninText,Ueki1,Eliashberg}
\begin{equation}
\rho = - i 2\pi e N (0) k_{\rm B} T \sum_{n = - \infty}^\infty \langle {g}_1 \rangle_{\rm F} - 2 e^2 N (0) \Phi \label{rho},
\end{equation}
where $\Phi$ is the static scalar potential.
Let us apply ${\bm \nabla}$ in Eq.\ (\ref{rho})
and use Maxwell's equations $\rho=\epsilon_0 {\bm \nabla} \cdot {\bm E}$ 
 and ${\bm\nabla}\times{\bm E}={\bm 0}$ with $\epsilon_0$ denoting the  vaccum permittivity.
We thereby obtain
\begin{equation}
- \lambda_{\rm TF}^2 {\bm \nabla}^2 {\bm E} + {\bm E} = i  \frac{\pi k_{\rm B} T}{e} \sum_{n = - \infty}^\infty \left\langle {\bm \nabla}g_1 \right\rangle_{\rm F}, 
\label{EEq}
\end{equation}
where $\lambda_{\rm TF} \!\equiv\! \sqrt{\epsilon_0/2 e^2 N (0)}$ is the Thomas--Fermi screening length. 
This equation enables us to calculate the electric field and charge density microscopically. 
Equations (\ref{nablag1}) and (\ref{EEq}) indicate that we can consider each of the PPG and Lorentz forces
independently to calculate the net charge using the principle of superposition to the order we are using.
Hence, we will study the two forces independently below.

We solved Eqs.\ (\ref{Ei}), (\ref{nablag1}), and (\ref{EEq})  numerically for an isolated vortex of an $s$-wave type-II superconductor with a cylindrical Fermi surface.
The magnetic field is directed along the side of the cylinder, and we choose the coordinate system where the vortex center is located on the $z$ axis.
To start with, we solved the standard Eilenberger equations (\ref{Ei1})-(\ref{Ei3}) self-consistently for the isolated vortex following the procedure described 
in Ref.\ \citen{KitaText}.
The solution was substituted into the right-hand side of Eq. (\ref{nablag1}), which was solved using the standard Runge--Kutta method. 
Substituting the solution of Eq.\ (\ref{nablag1}) into Eq.\ (\ref{EEq}),
and solving Eq. (\ref{EEq}) numerically, 
we obtained the electric field and charge density. 
The results presented below are for $\lambda_{\rm TF} = 0.01\xi_0$ and $\delta = 0.01$.

\begin{figure}[t]
        \begin{center}
                \includegraphics[width=0.9\linewidth]{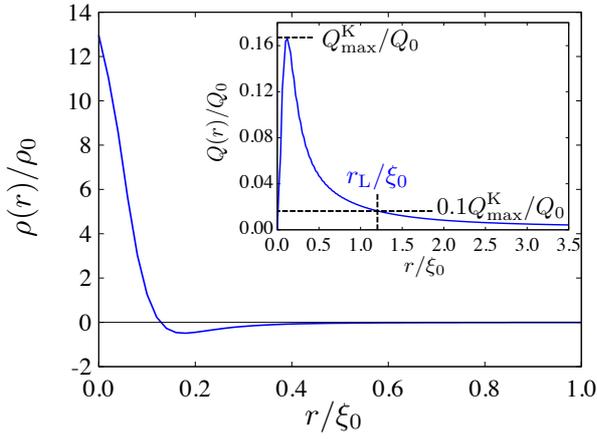}
                \end{center}
\caption{
(Color Online) Charge redistribution $\rho (r)$ due to the PPG force in units of $\rho_0 \equiv \epsilon_0 \Delta_0 /|e|\xi_0^2$ over $r \leq \xi_0$ at $T = 0.2 T_{\rm c}$, and accumulated charge $Q(r)$ in units of $Q_0 \equiv \epsilon_0 \Delta_0 /|e|$ over $r \leq 3.5\xi_0$.
}
\label{fig1}
\end{figure}
\begin{figure}[t]
        \begin{center}
                \includegraphics[width=0.9\linewidth]{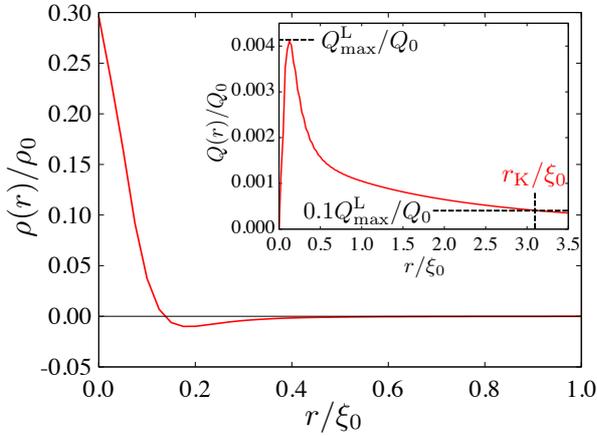}
                \end{center}
\caption{
(Color Online)  Charge redistribution $\rho (r)$ due to the Lorentz force in units of $\rho_0$ over $r \leq \xi_0$ at $T = 0.2 T_{\rm c}$, and accumulated charge $Q(r)$ in units of $Q_0$ over $r \leq 3.5\xi_0$.
}
\label{fig2}
\end{figure}

Figures \ref{fig1} and \ref{fig2} plot the radial dependence of the charge density in the core region due to 
the PPG and Lorentz forces,
respectively,
for $\lambda_0 = 5\xi_0$ at $T = 0.2 T_{\rm c}$.
The insets show the accumulated charge per unit length along the $z$ direction defined by
 \begin{align}
 Q(r)=2\pi\int^r_0\rho(r_1)r_1dr_1 .
 \label{Q(r)}
 \end{align}
The charge density and accumulated charge are normalized 
by the $\rho_0 \equiv \epsilon_0 \Delta_0 /|e|\xi_0^2$ and $Q_0 \equiv \epsilon_0 \Delta_0 /|e|$, respectively,
where $\lambda_0\equiv \sqrt{{\hbar}/{\mu_0 \xi_0|e|^2 N(0)\Delta_0 v_{\rm F}}}$ is the London penetration depth at zero temperature, and $T_{\rm c}$ denotes the superconducting transition temperature in a zero magnetic field. 
Quantities $Q_{\rm max}^{\rm K}$ and $Q_{\rm max}^{\rm L}$ are
the peak values of $Q(r)$ due to the PPG and Lorentz forces, respectively,
and $r_{\rm K}$ and $r_{\rm L}$ denote the distances from the core center at which
$Q(r_{\rm K})= 0.1Q_{\rm max}^{\rm K}$ and  $Q(r_{\rm L})= 0.1Q_{\rm max}^{\rm L}$ are satisfied, respectively.
We observe that the PPG force causes charge accumulation at the core center whose magnitude is about 50 times larger than that caused by the Lorentz force. 
This charge accumulation by the PPG force is concentrated in the core region of $r\lesssim \xi_0$, 
as expected naturally by noting that it is caused by the reduction in the pair potential. Indeed, the charge neutrality is
recovered for $r\sim \xi_0$.
However, 
the charge accumulation by the Lorentz force is smaller at the core center but extends far outside the core
over $r\lesssim \lambda_{\rm L}$,
where $\lambda_{\rm L}$ denotes the London penetration depth 
defined by $\lambda_{\rm L}(T)=\lambda_0[1-Y(T)]^{1/2}$ 
in terms of the Yosida function $Y(T)$.\cite{Kita09,KitaText,Yosida}.  
This charge extension becomes broader as $\lambda_{\rm L}$ increases  for $T\gtrsim 0.5T_{\rm c}$.

\begin{figure}[t]
        \begin{center}
                \includegraphics[width=0.9\linewidth]{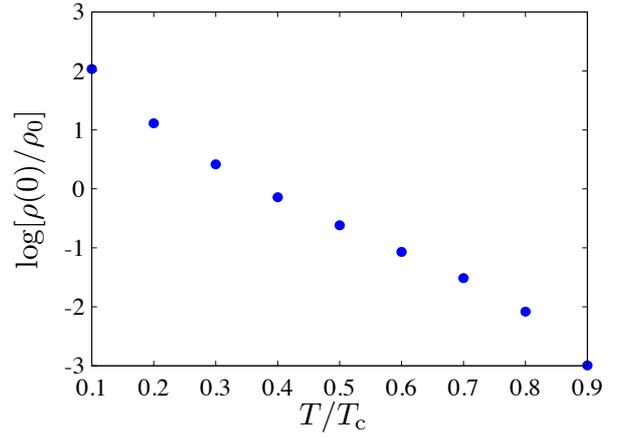}
                \end{center}
\caption{
(Color Online) Normalized charge density due to the PPG force at the vortex center as a function of temperature calculated for $\lambda_0=5\xi_0$. 
}
\label{fig3}
\end{figure}
\begin{figure}[t]
        \begin{center}
                \includegraphics[width=0.9\linewidth]{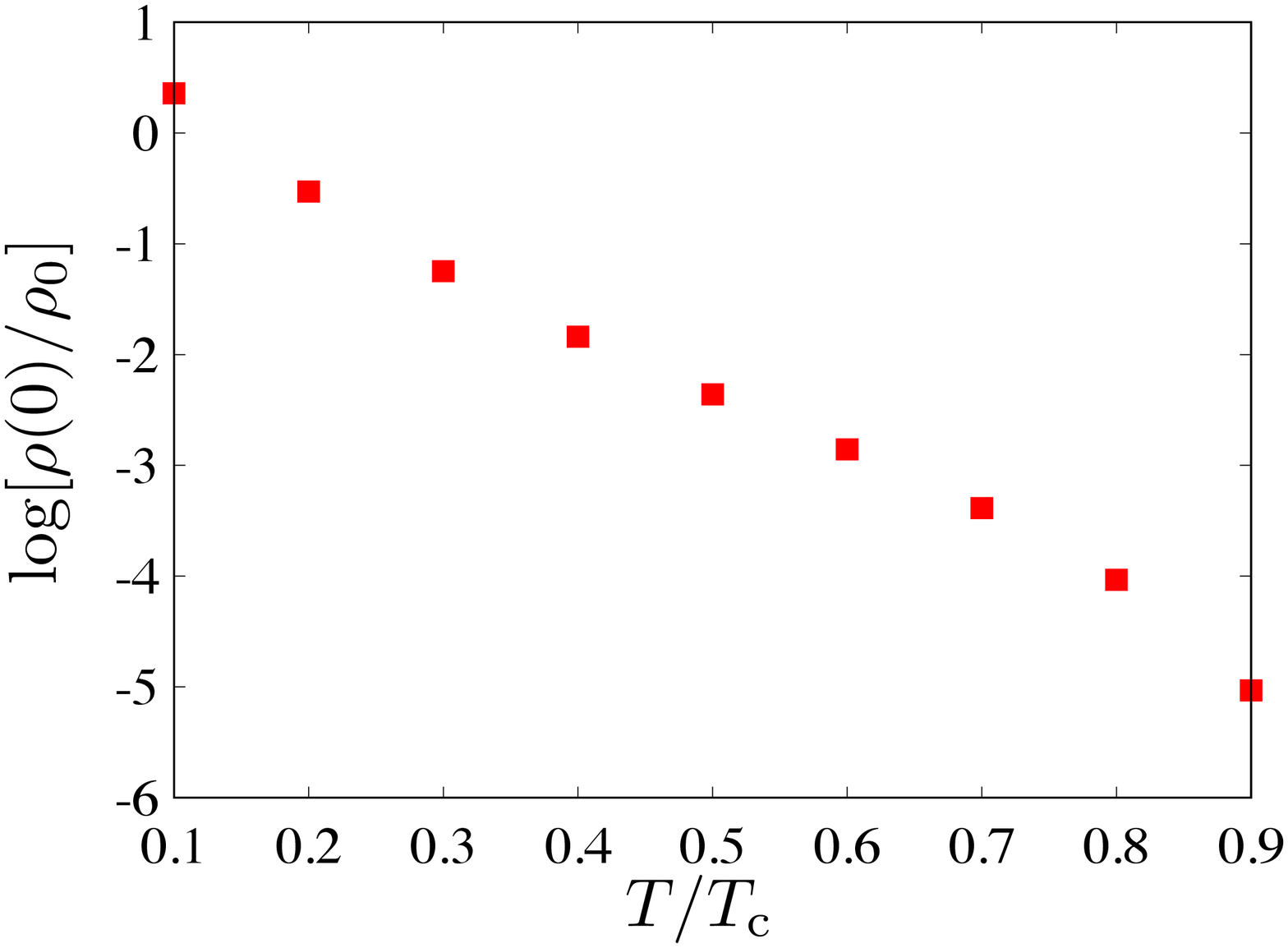}
                \end{center}
\caption{
(Color Online) Normalized charge density due to the Lorentz force at the vortex center as a function of temperature calculated for $\lambda_0=5\xi_0$. 
}
\label{fig4}
\end{figure}
\begin{figure}[t]
        \begin{center}
                \includegraphics[width=0.9\linewidth]{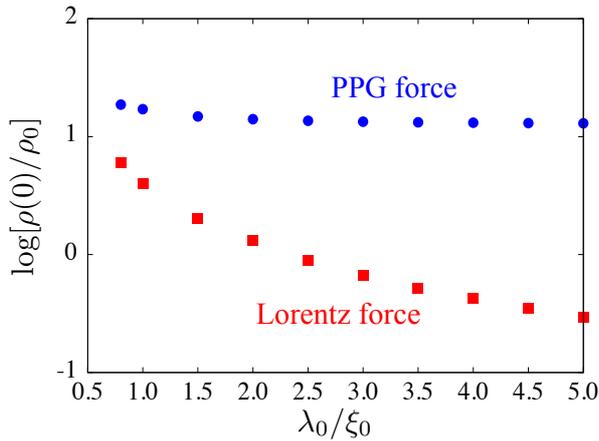}
                \end{center}
\caption{
(Color Online) Normalized charge density due to the PPG force (blue points) and Lorentz force (red points) at the vortex center as a function of $\lambda_0$ calculated for $T=0.2T_{\rm c}$.
}
\label{fig5}
\end{figure}

Figures \ref{fig3} and \ref{fig4} show the temperature dependence of the logarithm of charge density 
at the vortex center due to the PPG and Lorentz forces, respectively, 
calculated for $\lambda_0=5\xi_0$. 
In both cases, $\rho(0)$ grows significantly as the temperature is decreased,
which can be attributed to the decrease in $\lambda_{\rm L}$ and also to the core shrinkage due to the Kramer--Pesch effect\cite{KP}
that causes an enhancement in the slope of ${\Delta(r)}$ at $r=0$.

Figure \ref{fig5} shows the $\lambda_0/\xi_0$ dependence of the logarithm of the charge density at the core center due to the PPG and Lorentz forces, respectively,
calculated at $T=0.2T_{\rm c}$.
In both cases, $\rho(0)$ increases as  $\lambda_0/\xi_0$ is decreased.
The contribution of the Lorentz force is negligible compared with that of the PPG force for $\lambda_0\gtrsim 2\xi_0$,
but becomes substantial when $\lambda_0\sim \xi_0$.
In this context, it is worth noting that $\rho(0)$ has been shown to have a strong magnetic-field dependence with a peak structure and can be enhanced significantly 
over the value of an isolated vortex as the magnetic field is increased.\cite{Kohno1,Kohno2}
Hence, it is not appropriate to conclude that the Lorentz-force contribution is negligible in all cases.

In summary, we have solved the augmented quasiclassical equations of superconductivity 
with the PPG and Lorentz forces
in the Matsubara formalism for an isolated  vortex of an $s$-wave type-II superconductor with an isotropic Fermi surface.
Our results reveal that the PPG force contributes dominantly to charging in the core region of an isolated vortex
over a wide parameter range. 
The charge accumulation due to the PPG force satisfies the charge-neutrality condition within the core region, 
whereas that  caused by the the Lorentz force extends outside the core up to the distance of the order of the London penetration depth.
It should also be pointed out that the two contributions are additive at the core center for the present model
with an isotropic Fermi surface and energy gap.
In this context, our previous studies\cite{Kohno1,Kohno2} have revealed that the charge accumulation originating from the Lorentz force
is enhanced substantially in finite magnetic fields.
This suggests that the Lorentz-force contribution may become dominant over the PPG force contribution in finite magnetic fields.
In addition, it is not clear whether the two contributions still work additively in the presence of Fermi-surface and gap anisotropies.
These issues remain to be clarified in the future.

\end{document}